\documentclass[12pt]{article} 

\usepackage{amsmath, amssymb, amsthm, verbatim} 

\usepackage[all]{xy}

\newcommand {\C } {\mathbb{C}} 

\newtheorem*{thma}{Theorem}

\newcommand {\p } {\mathbb{P}} 

\newcommand {\R } {\mathbb{R}} 

\newcommand {\T} {T^* S^3} 

\newcommand {\op} {$\hat Y_{\alpha,b}$ }

\newcommand {\cl} {$Y_{z_1,z_2}$} 

\newcommand {\OO} {$\mathcal {O}(-1) \oplus \mathcal {O} (-1) \rightarrow \mathbb{P}^1$}

\newcommand {\Z} {\mathbb{Z}}

\begin{document} 

\title{\bf{Computations on $B$-model geometric transitions}} 

\author{Brian Forbes \\ \\ \it Division of Mathematics, Graduate School of Science \\ \it Hokkaido University \\ \it Kita-ku, Sapporo, 060-0810, Japan \\ \it brian@math.sci.hokudai.ac.jp}

\date{July, 2004} 

\maketitle

\begin {abstract} 
We study geometric transitions on Calabi- Yau manifolds from the perspective of the $B$ model. Looking toward physically motivated predictions, it is shown that the traditional conifold transition is too simple a case to yield meaningful results. The mathematics of a nontrivial example \cite{AV} is worked out carefully, and the expected equivalence is demonstrated.

\end {abstract}

\section {Introduction.}

Dualities in physics have led to many surprising connections in mathematics. In mirror symmetry \cite{CK}, we find that the variation of Hodge structure on a family of Calabi-Yau manifolds gives enumerative information on curve number for a completely different Calabi-Yau family. More recently \cite{OV},\cite{MV}, \cite{LLZ}, it has been shown that Chern-Simons computations on $T^*S^3$ can be used to compute disc numbers, as well as Hodge integrals, on \OO.    

The equivalences we look for depend on which string theories are exchanged under the given physical duality. For instance, in the case of mirror symmetry, type $A$ and $B$ string theories are exchanged; we can have either closed or open $A$ (resp. $B$) string theory on each side \cite{LM},\cite{BF}. In the case that both sides are closed, we are exchanging the complexified K\"{a}hler modulus on the $A$ family with the complex modulus on the $B$ family; if both sides are open, then in addition we must consider the position of Lagrangian submanifolds on the $A$ side and holomorphic submanifolds on the $B$ side. 

For the geometric transition $(\T) \rightarrow$(\OO), in which the $S^3$ of $\T$ shrinks and is replaced by a blown up $\p^1$, the basic duality is that open $A$ type strings on $\T$ are supposed to be equivalent to closed $A$ type strings on \OO. Taking the mirror of this, if $Y_{z_1}$ is the mirror of \OO and $\hat Y_s$ is the mirror of $\T$, then we expect an equality between open $B$ strings on $\hat Y_s$ and closed $B$ strings on $Y_{z_1}.$

The purpose of this paper is to clarify the mathematics of such proposed dualities \cite{AV}, taking the above case as a starting point. While the computations are motivated by physics, the calculations themselves are purely mathematical and require no physical background. 

Section 2 reviews the $A$ and $B$ model of the conifold transition, and points out problems in this case. Section $3.1-3.2$ works out the geometric transition for a more complex case, and $3.3$ describes the predicted duality. Sections $3.4-3.5$ compute the functions that should be equal under the duality, and $3.6$ demonstrates equality.

\bigskip

$\bf{Acknowledgements.}$ 
I would like to thank Masaki Shigemori for helpful discussions, and Shinobu Hosono and Simeon Hellerman for comments. Finally, I thank Kefeng Liu and the Center of Mathematical Sciences at Zhejiang University for their hospitality while this work was completed.

\bigskip

\section {Review of the Conifold Transition.} 

The aim of this section is to provide exposition on the simplest 
possible example of equivalences through geometric transitions. Here 
we will show that the basic conifold transition is not sufficient to yield mathematical predictions for the $A$ or $B$ model 
transition. 

\subsection {A-model perspective.} 

Before giving the details of the spaces involved in the $B$ model, 
it is helpful to first review the original formulation of the 
conifold transition, namely, when the spaces on both sides of the 
transition are considered in the $A$ model case. First, we will review the geometry of the conifold transition, and afterwards describe what mathematical equalities are expected to follow from the physics.

We begin with the space $\hat X_a = \T ,$ which can be given as a hypersurface 
$$\{(w_1, \dots , w_4) \in \C^4 : w_1w_2+w_3w_4+a=0\}.$$ 
Here, let $a \in \R_{\ge0},$ so that 
$$S^3=\T \cap \{w_2=-\bar{w_1}, w_4=-\bar{w_3}\}.$$
 We have that $S^3$ is trivially a Lagrangian submanifold of $\T.$ Note that this space develops a singularity as $a\rightarrow 0,$ and furthermore that near the singular point $w_1=\dots =w_4=0$ in $\hat X_0,$ the topology is that of a cone.  

To complete the transition, we can blow up the singularity of $\hat X_0.$ Let $[x_0,x_1]$ be homogeneous coordinates on $\p^1,$ and define $X_r=$
$$
\{(w_1,\dots , w_4, [x_0, x_1])\in \C^4 \times \p^1 : w_1x_0-w_3x_1=0, w_2x_1+w_4x_0=0\}.
$$
Above, $r$ is the real K\"{a}hler parameter depending on the choice of K\"{a}hler class corresponding to the exceptional $\p^1.$ It is known that there is a diffeomorphism $X_r \cong$ \OO \cite{GR}.

To summarize, then, the geometric transition just considered is
\begin{equation}
\label{Atrans}
\hat X_a \stackrel{a\rightarrow 0}{\longrightarrow} \hat X_0 \stackrel{blowup}{\longrightarrow} X_r.
\end{equation}

Now we will briefly discuss the physical duality, and what this means for mathematicians. It is expected that closed $A$ model topological strings on \OO are equivalent to open $A$ model strings on $\T$. Now, for example, closed string mirror symmetry postulates a local isomorphism between the complexified K\"{a}hler moduli space of a family of $A$ model Calabi-Yaus and the complex moduli space of a different family of Calabi-Yaus in the $B$ model. Here, there is a similar type of equivalence, except that there are $A$ model strings on both sides of the transition. This means that we must match a function of the closed string $A$ model moduli space (namely, the complexified K\"{a}hler moduli space of \OO) with a function of the open string $A$ model moduli space on $\T$.

The $A$ model moduli space, generally speaking, consists of the complexified K\"{a}hler moduli together with the location of Lagrangian $3$ cycles in the Calabi-Yau. So, to see what physics has to say about mathematics, we would like to match the Gromov-Witten potential, which is a function of the complexified K\"{a}hler moduli of \OO, with a certain function (called the superpotential) defined on the open $A$ model moduli space of $\T.$

 For our case $\T$ above, we can already see the problem with this approach; there is no K\"{a}hler modulus on this space, and the Lagrangian $S^3$ is rigid inside $\T$. Thus, there is nothing for the superpotential to be a function of as there are no $A$ model moduli. Indeed, it is the case that the physical superpotential on $\T$ is not a function of any geometric quantity.

\subsection{The $B$ model conifold transition.}

Next, we will carry out the reverse transition on the mirror of the above construction (\ref{Atrans}), following \cite{B}, \cite{AV}. Recall \cite{AV2}, \cite{V} that the mirror of the space $X_r$ is given by $Y_{z_1}=$
\\
$$
\{(x,z,y_3,y_4)\in \C^2\times (\C^*)^2: xz+1+y_3+y_4+z_1y_3y_4=0\}.
$$
\\
To see how to reverse the conifold transition, beginning with the space $Y_{z_1},$ note that if $t=r+i\theta$ is the complexified K\"{a}hler modulus on $X_r =$\OO, then we have $z_1=e^{-t}.$ Thus, since $X_r \rightarrow \hat X_0$ as $r \rightarrow 0,$ we should let $z_1 \rightarrow 1$ on $Y_{z_1}.$ The result is $\hat Y_1=$
\\
$$
\{(x,z,y_3,y_4)\in \C^2\times (\C^*)^2: xz+(1+y_3)(1+y_4)=0\},
$$
\\
which has a singularity where $x=z=1+y_3=1+y_4=0.$ As above, this can be blown up to get $\hat Y_s=$
$$
\{(x,z,y_3,y_4,[x_0,x_1])\in \C^2\times (\C^*)^2 \times \p^1:  
$$
$$xx_0-(1+y_3)x_1=0,zx_1+(1+y_4)x_0=0\}.
$$
Again, $[x_0,x_1]$ are homogeneous coordinates on the exceptional $\p^1$ and $s$ is determined from the real K\"{a}hler parameter on $\p^1$.

With these considerations, we have the extended diagram

\begin{displaymath}
\xymatrix{ \hat X_a \ar[r] & \hat X_0 \ar[r] & X_r  \ar[d] \\ 
          \hat Y_s  & \hat Y_1 \ar[l]  & Y_{z_1} \ar[l] }
\end{displaymath}

The vertical arrow represents mirror symmetry, and the other arrows are as given by the various geometric transitions already described. It is believed that $\hat Y_s$ is the mirror of $\hat X_a$ \cite{B}.

Now, let's again discuss physical predictions, this time from the perspective of the $B$ model transition. Naturally, we should again see no mathematical conjecture emerging from physics, as there was none from the $A$ model transition.

The closed string modulus of $Y_{z_1}$ in the $B$ model is the complex structure modulus $z_1$. The relevant mathematical quantity on the moduli space of $Y_{z_1}$ that we would like to compare to a quantity on the moduli space of $\hat Y_s$ is the period
$$
W(z_1) = \frac{1}{2}(log(z_1))^2+\sum_{n>0}\frac{z_1^n}{n^2}.
$$
Note that, if we perform the indefinite integral
$$
\int W(z_1) \frac{dz_1}{z_1},
$$
the second term becomes $\sum_{n>0} z_1^n/n^3.$ This is as expected, because after accounting for the multiple cover formula, this predicts the existence of a single holomorphic sphere in the mirror geometry. Clearly this is the case for \OO.

(There is a peculiarity in the Picard- Fuchs system on the mirror of \OO \ in that it does not annihilate $W(z_1).$ This will be addressed in a forthcoming paper.)

To make a meaningful comparison between the function $W(z_1)$ and the function on the moduli space of $\hat Y_s,$ we are supposed to set $z_1=1$ in the above \cite{AV}\cite{B}. This corresponds to the limit $z_1\rightarrow 1$ that was passed through in the construction of $\hat Y_s.$ Then we get $W(z_1=1)=const$.

Next, we move to the moduli space of $\hat Y_s.$ As this is in the open $B$ model, the moduli are the complex structure modulus and the position of the exceptional $\p^1$. Now, from \cite{W} we have that the quantity on the moduli space of $\hat Y_s$ that should match $W$ above is the superpotential
$$
\hat W(v)= \int_{\Gamma(v)} \Omega_s.
$$
Here $\Gamma(v)$ is a 3-manifold such that $\partial \Gamma(v)=\p^1 - \p^1(v)$(that is, a 1 parameter family of exceptional $\p^1$'s) and $\Omega_s$ is a holomorphic $(3,0)$ form on $\hat Y_s.$ Also, $\p^1$ is some fixed representative in the homology class of the exceptional $\p^1$. As the $\p^1$ is rigid, $\hat W$ is zero as expected; thus, we need a more nontrivial geometry on which to test $B$ model geometric transitions. Still, we find some correspondence in the sense that $ \hat W(v)$ and $W(z_1=1)$ are both constants.

\section{Review of a generalized conifold transition.}

Next, we would like to imbed the above transition in a more complex case, as in \cite{AV}. Note that, for a geometric transition to make sense in physics, there must always be a $\p^1$ in the total space (say $X$) such that $\mathcal N_{\p^1/X}\cong \mathcal O (-1) \oplus \mathcal O (-1)$. In the case studied below, there are now 2 $\p^1$'s, with respective normal bundles $ \mathcal O (-1) \oplus \mathcal O (-1)$ and $ \mathcal O  \oplus \mathcal O (-2)$. We will shrink the $\p^1$ with normal bundle $ \mathcal O (-1) \oplus \mathcal O (-1)$ and deform the resulting singularity so that an $S^3$ is in the place of the $\p^1$ (this is actually the reverse direction of the transition shown in the sequel).

The equivalence of $A$ model functions on the geometric transition has been worked out in \cite{DFG}, and some physical calculations on the $B$ model have been done in \cite{AV}. The aim of the work in this paper is to put such calculations on clear mathematical footing. 

\subsection{Setup of the $A$-model.}

For consistency with the previous section, we derive the same chain of transitions for this new case. Set $\hat X_{r,a}=$
\begin{equation}
\label{spaceA}
\{(w_1,w_2,w_3,[x_0,x_1]\in \C^3\times \p^1 : w_1x_0+w_2w_3x_1+ax_1=0\}.
\end{equation}
Again, $a\in \R_{\ge 0},$ and $r$ is the real K\"{a}hler parameter corresponding to the $\p^1$.  On this space, we have two coordinate patches corresponding to the hemispheres of the $\p^1$; the local equations in each are given by
\\
$$
\hat{\mathcal U}_1(r,a)=\{w_1x+w_2w_3+a=0\}, \ \ \ \hat{\mathcal U}_2(r,a)= \{w_1+w_2w_3x'+ax'=0\}
$$
\\
where $x=x_0/x_1$ and the transition function is $x=1/x'.$
In the first coordinate patch, as $a \rightarrow 0$ our space can be seen to develop the same singularity as we had in section 2. Define 
\\
$$
\hat X_{r,0} = \hat{\mathcal U}_1(r,0) \cup \hat{\mathcal U}_2(r,0).
$$
\\
Then we blow the first coordinate patch up along $w_1=w_2=0$ to get 

$$
\hat{\mathcal V}_1(r,s) = \{w_1u_0-w_2u_1=0, xu_1+w_3u_0=0\}
$$
\\
with $[u_0,u_1]$ homogeneous coordinates on $\p^1$ and $s$ determined by the exceptional $\p^1$. Finally, set
\\
$$
X_{r,s} = \hat{\mathcal V}_1(r,s) \cup \hat{\mathcal U}_2(r,0).
$$
\\
It is known \cite{DFG}\cite{AV} that this space contains 2 $\p^1$'s, $C_1$ and $C_2$, such that ${\mathcal N}_{C_1/ X_{r,s}} \cong \mathcal O \oplus \mathcal O(-2)$ and ${\mathcal N}_{C_2/ X_{r,s}} \cong \mathcal O(-1) \oplus \mathcal O(-1).$

Then, the equivalence of the relevant $A$ model quantities on $X_{r,s}$ and $\hat X_{r,a}$ has already been shown \cite{DFG}. Thus, we move directly to the mirror of this transition.

\subsection{The $B$ model.}

We have that the mirror of the space $X_{r,s}$ is given by the hypersurface $Y_{z_1,z_2}=$
\begin{equation}
\label{closedB}
\{(x,z,y_1,y_2)\in \C^2\times (\C^*)^2 :  xz+1+y_1+y_2+z_1y_1^{-1}+z_1z_2y_1^{-1}y_2=0\}.
\end{equation}
The relationship between $(r,s)$ and the complex variables $(z_1,z_2)$ is: if $t_1=r+i\theta_1, \ t_2=s+i\theta_2$ are the complexified K\"{a}hler parameters, then $z_i=e^{-t_i}, \ i=1,2$. 

From the previous section, recall that $s$ corresponds to the size of the curve with normal bundle ${\mathcal N}_{C_2/ X_{r,s}} \cong \mathcal O(-1) \oplus \mathcal O(-1)$. This means that we expect to take a limit $z_2\rightarrow 1$ in order to pass the space $Y_{z_1,z_2}$ through the reverse conifold transition. However, there is a physical subtlety at this point which states that there should be corrections to the variable $z_2$, and thus $z_2\rightarrow 1 $ is not quite the right limit. This will be clarified in the section on the Picard- Fuchs operators; for now, in order to exhibit the singularity of the intermediate space $Y_{z_1,1}$, we simply perform a change of variables. This follows \cite{AV}.

Then, make the definitions
\begin{equation}
\label{changevar}
y_i=\frac{v_i}{1+\alpha},  \ z_1=\frac{\alpha}{(1+\alpha)^2}, \ z_2=\beta(1+\alpha)
\end{equation}
for $i=1,2$, with $\alpha, \beta \in \C^*$ and $|\alpha|<1$.

With respect to these coordinates, we arrive at the transformed equation for $Y_{z_1,z_2}$, which is $Y_{\alpha,\beta}=$
\begin{equation}
\label{closedB}
\{(x,z',v_1,v_2)\in \C^2\times (\C^*)^2 :  xz'+1+\alpha+v_1+v_2+\alpha v_1^{-1}+\alpha \beta v_1^{-1}v_2=0\},
\end{equation}
with $z'=z(1+\alpha)$. Then, taking $\beta \rightarrow 1$, we get the singular space $\hat Y_{\alpha,1}=$
\\
\begin{equation}
\label{singularB}
\{(x,z',v_1,v_2)\in \C^2 \times (\C^*)^2 :  xz'+(1+v_1+v_2)(1+\alpha v_1^{-1})=0\},
\end{equation}
\\
and then the natural thing is to blow up the singularity like we did earlier; if $[x_0,x_1]$ are homogeneous coordinates on $\p^1$, the result is $\hat Y_{\alpha,b}=$
\\
\begin{equation}
\label{openB}
\{xx_0-(1+v_2+v_2)x_1=z'x_1+(1+\alpha v_1^{-1})x_0=0\}
\end{equation}
\\
which lives in $\C^2 \times (\C^*)^2 \times \p^1,$ and $b$ is determined by the exceptional $\p^1$.

After all these considerations, we have a diagram for this more complex case:

\begin{displaymath}
\xymatrix{ \hat X_{r,a} \ar[r] & \hat X_{r,0} \ar[r] & X_{r,s}  \ar[d] \\ 
          \hat Y_{\alpha,b}  & \hat Y_{\alpha, 1} \ar[l]  & Y_{z_1,z_2}=Y_{\alpha, \beta} \ar[l] }
\end{displaymath}
\\
This summarizes all transitions involved. Next, we would like to clarify the predicted mathematical equivalences.

\subsection{Expected dualities on the $B$ model transition.}

Similarly to the usual conifold transition case, the spaces on which we hope to match functions are \op and \cl. More precisely, from physics it is expected that the relevant quantity for comparison coming from \cl \ is a particular period of \cl, which is a function of $z_1, z_2$; this is because we are working with closed strings on \cl. In order to determine the period, it suffices to write down the Picard- Fuchs system for \cl, and use the Frobenius method to generate the appropriate function for comparison.

In fact, there is a bit more to it than just this, as was observed in \cite{AV},\cite{B}. Across the geometric transition, we are taking a $\p^1$ to zero size, but according to physics we expect that there are corrections to the size of the $\p^1.$ This will be discussed in detail in the sequel.

Next, we must identify the corresponding function on the moduli space of \op. As before, this moduli space consists of the complex structure moduli of \op together with the position of the exceptional $\p^1.$ Now in this case, the $\p^1$ actually moves in a 1 parameter family; therefore the earlier mentioned integral 
\\
$$
\hat W(v)= \int_{\Gamma(v)} \Omega_{\alpha, b}
$$
\\
will yield a nontrivial (and 1 parameter) solution for this case. The object, then, is to match the appropriately identified period on \cl \ with the function $\hat W(v)$ defined on the open string moduli space of \op. The next sections will carefully derive these relations.

\subsection{Periods on \cl.}

While periods on noncompact Calabi-Yaus have yet to be put on completely rigorous grounds, their heuristics for some cases have been explained \cite{CKYZ}, and a recent paper of Hosono \cite{H} clarified the meaning of noncompact period integrals.

We are working with the space $Y_{z_1,z_2}=$
\\
$$
\{(x,z,y_1,y_2)\in \C^2\times (\C^*)^2 :  xz+1+y_1+y_2+z_1y_1^{-1}+z_1z_2y_1^{-1}y_2= f= 0\}.
$$
\\
Then \cite{H} the period integrals of \cl \ are defined to be
\\
\begin{equation}
\label{periods}
W_{\Gamma}(z_1,z_2)=\int_{\Gamma}\frac{dxdz\frac{dy_1}{y_1}\frac{dy_2}{y_2}}{xz+1+y_1+y_2+z_1y_1^{-1}+z_1z_2y_1^{-1}y_2}
\end{equation}
\\
for $\Gamma \in H_4(\C^2\times (\C^*)^2 -f, \Z).$

In this form, it's a bit difficult to see what the Picard- Fuchs system annihilating the $W_{\Gamma}(z_1,z_2)$ ought to be. So, we perform a standard trick of enlarging the moduli space and taking a quotient at the end. This is equivalent to adding additional GKZ operators to the Picard- Fuchs system. 

Then, the integrals we would like to consider are
\\
$$
\tilde W_{\Gamma}(a_0,\dots,a_4)=\int_{\Gamma}\frac{dxdz\frac{dy_1}{y_1}\frac{dy_2}{y_2}}{xz+a_0+a_1y_1+a_2y_2+a_3y_1^{-1}+a_4y_1^{-1}y_2}.
$$
\\
From this, it is easy to produce operators that annihilate the $\tilde W_{\Gamma}(a_0,\dots,a_4);$ these are
\\
$$
{\mathcal L}_1=\partial_{a_1}\partial_{a_3}-\partial_{a_0}^2, \
{\mathcal L}_2=\partial_{a_0}\partial_{a_2}-\partial_{a_1}\partial_{a_4}. 
$$
\\
After this, we can use standard techniques \cite{CK} \cite{CKYZ} as follows. The operators ${\mathcal L}_1, \ {\mathcal L}_2$ determine two canonical variables
\\
$$
z_1=\frac{a_1a_3}{a_0^2}, \ z_2=\frac{a_0a_2}{a_1a_4},  .
$$
\\
In terms of these variables, ${\mathcal L}_1, \ {\mathcal L}_2$ can be rewritten as follows. Set $\theta_i = z_i\frac{\partial}{\partial z_i}, \ i=1,2.$ The new operators read
\\
\begin{equation}
\label{PFops}
{\mathcal D}_1 = \theta_1(\theta_1-\theta_2)-z_1(2\theta_1-\theta_2)(1+2\theta_1-\theta_2), 
\end{equation}
$$
{\mathcal D}_2 = (2\theta_1-\theta_2)\theta_2-z_2(\theta_1-\theta_2)\theta_2.
$$
\\
The solutions of these will be the periods. All solutions are generated, via the Frobenius method, from the function
\\
$$
\omega_0(z,\rho) = \sum_{n\ge 0}c(n,\rho)z_1^{n_1+\rho_1}z_2^{n_2+\rho_2
},
$$
\\
where
\\
$$
c(n,\rho)\ =
$$
$$ 
[\Gamma(1-2n_1-2\rho_1+n_2+\rho_2)\Gamma(1+n_1+\rho_1-n_2-\rho_2)*
$$
$$
\Gamma(1+n_2+\rho_2)\Gamma(1+n_1+\rho_1)\Gamma(1-n_2-\rho_2)]^{-1}.
$$
\\
Then, we find two logarithmic solutions, which are 
\\
\begin{equation}
\label{mirrormap}
t_1(z)=log(z_1)+2\sum_{n_1 > 0} \frac{(2n_1-1)!}{(n_1!)^2}z_1^{n_1}, \
t_2(z)=log(z_2)-\sum_{n_1} \frac{(2n_1-1)!}{(n_1!)^2}z_1^{n_1}.
\end{equation}
\\
At this point, it is possible to explain the need for the change of variables in the $B$ model given in section 3.2. It is most natural to just let $z_2\rightarrow 1$ above, but from physics there are corrections to the volume of the $\p^1$. The mirror map $t_1,t_2$ represents the corrected volume of the $\p^1$'s; therefore, by changing variables to $t_1,t_2$ and taking $t_2$ to zero, the $B$ model geometry will pass through the looked-for singularity. The reason this problem did not emerge for the usual conifold transition is that, for that example, the mirror map is trivial, and hence no correction is needed. 

Next, we give the double logarithmic solution of the PF system (modulo the log terms):
\\
\begin{equation}
\label{closedsuper}
W(z_1,z_2)= \sum_{n_2>n_1\ge0}\frac{(-1)^{n_1}(n_2-n_1-1)!}{(n_2-2n_1)!n_1!n_2}z_1^{n_1}z_2^{n_2}.
\end{equation}
\\

Now, let us discuss some general features of the solutions of the system (\ref{PFops}). We have that $t_1, \ t_2$ and $W$ are solutions, but how can we determine which one (if any) of these is the right one for comparison with a function defined on the moduli space of \op? Recall \cite{CKYZ} \cite{H} that the solutions generated by the Frobenius method of such a noncompact system of Picard- Fuchs equations can be arranged in a period vector
\\
$$
\Pi(z_1,z_2)=(1,\ t_1, \ t_2, \ \frac{\partial {\mathcal F}}{\partial t_1}, \  \frac{\partial {\mathcal F}}{\partial t_2},\dots)
$$
\\
Above, $t_1, \ t_2$ are the mirror map (and also the logarithmic solutions), and $\mathcal F$ is the prepotential, which gives enumerative predictions for curve counting in the mirror geometry (after the insertion of the inverse mirror map and the multiple cover formula are included). Further, \ $\partial {\mathcal F}/\partial t_i$ are the double logarithmic solutions.

  With this notation, the equation (\ref{closedsuper}) becomes
\\
$$
W(z_1,z_2)=\frac{\partial {\mathcal F}}{\partial t_2}(z_1,z_2).
$$
\\
The reason that this has been singled out for comparison is as follows. Of the two double logarithmic solutions of the system, one ($\partial {\mathcal F}/ \partial t_1$) will not be well defined, because the imbedded $\p^1$ in the mirror geometry corresponding to the variable $t_1$ moves in an (unbounded) 1-parameter family; thus we do not expect sensible enumerative predictions from that particular double log solution. That leaves only one for consideration, which we identify with $W$ (as in \cite{V}, etc).

This concludes the period integral computation on \cl. We now turn to the (open string) calculation on \op. 

\subsection{The open string superpotential on \op.}

We now refer back to the defining equations (\ref{openB}) of \op:
\\
$$
\{xx_0-(1+v_2+v_2)x_1=z'x_1+(1+\alpha v_1^{-1})x_0=0\}.
$$
\\
Recall that $(x,z',v_1,v_1,[x_0,x_1])\in \C^2\times(\C^*)^2\times\p^1.$ Let $u=x_0/x_1$ for $x_1\ne0$, and $u'=x_1/x_0$ for $x_0\ne0$. Then \op has two coordinate patches, ${\mathcal W}_i= Y_{\alpha,b}\cap \{x_i\ne0\}, \ i=0,1$,  in which we get local equations:
\\
\begin{equation}
\label{patch0}
{\mathcal W}_0=\{x=(1+v_1+v_2)u, \ z'u=-(1+\alpha v_1^{-1})\},
\end{equation}
\begin{equation}
\label{patch1}
{\mathcal W}_1=\{xu'=1+v_1+v_2, \ z'=-(1+\alpha v_1^{-1})u'\}.
\end{equation}
\\
In order to compute the thing we would like to use to compare with the closed string caculation of the previous section, we need to find a holomorphic $(3,0)$ form on \op. 

From the blowup of section $3.1$, we get a projection map
\\
$$
\pi : Y_{\alpha,b} \longrightarrow Y_{\alpha,1},
$$
\\
so we can pull back a form on $Y_{\alpha,1}$ to each coordinate patch ${\mathcal W}_0, \ {\mathcal W}_1.$ For the form on $Y_{\alpha,1}$, use
\\
$$
Res\Big( \frac{dxdz'\frac{dv_1}{v_1}\frac{dv_2}{v_2}}{xz'+(1+v_1+v_2)(1+\alpha v_1^{-1})}\Big).
$$
\\
Then we will take
\\
$$
{\tilde \Omega}_0= \frac{dz'dv_1dv_2}{z'v_1v_2}, \ {\tilde \Omega}_1= \frac{dxdv_1dv_2}{xv_1v_2}.
$$
\\
There are two restricted projection mappings
\\
$$
\pi_i:{\mathcal W}_i\longrightarrow Y_{\alpha,1}, \ \pi_i=\pi|_{{\mathcal W}_i},\ i=0,1,
$$
\\ 
and then we can pull back the forms with these to get $\Omega_i=\pi_i^*{\tilde \Omega}_i$.

To do this properly, from the defining equations (\ref{patch0}),(\ref{patch1}) one can see that $(z',u, v_2)$ is a natural set of coordinates on ${\mathcal W}_0$ and similarly $(x,u',v_1)$ are coordinates on ${\mathcal W}_1.$ Then, from the two restriction maps $\pi_i$, we find
\\
$$
\Omega_0= \frac{dz'dudv_2}{(1+z'u)v_2}, \ \Omega_1=\frac{dxdu'dv_1}{(xu'-1-v_1)v_1}.
$$
\\
This gives a $(3,0)$ form for \op. 

Next, recall \cite{W} that the superpotential, which is the function we are trying to match, can be defined as an integral
\\
$$
{\hat W}_i(v) = \int_{\Gamma(v)}\Omega_i
$$
\\
where the $\Omega_i$ are as above and $\Gamma(v)$ is a 1 parameter family of $\p^1$'s; so, $\Gamma \cong \p^1 \times I$, where $I$ is a closed interval. More specifically, in the first (resp. second) patch, $v_1$ ($v_2$) will be the deformation parameter on the exceptional $\p^1$ with respect to a fixed representative in its homology class, and the other coordinates will be taken as $\p^1$ coordinates. We will integrate over each half of the $\p^1$ separately. 

First, work on ${\mathcal W}_0$. The integral then looks like
\\  
$$
\hat W_0(v_2^*)= \int_{\Gamma(v_2^*)}\Omega_0=\int_{v_2=\epsilon}^{v_2=v_2^*}\int_{\p^1_-}\frac{d{\bar u}dudv_2}{(1+{\bar u}u)v_2}
$$
\\
where we have identified $v_2$ as the coordinate parameterizing the family $\Gamma$, and also have set $z'=\bar u$ since the integration is over the $\p^1$. This integral gives 
\\
\begin{equation}
\label{superpotential0}
\hat W_0(v_2^*)=\int_{v_2=\epsilon}^{v_2=v_2^*}\int_{r=0}^{r=r_0}\int_{\theta=0}^{\theta=2\pi}\frac{rdrd\theta}{1+r^2}\frac{dv_2}{v_2}=\pi log(1+r_0^2)log(v_2^*) - c,
\end{equation}
\\
and here $c=\pi log(1+r_0^2)log(\epsilon)$. The choice $v_2=\epsilon$ represents a fixed element in the homology class of the exceptional $\p^1$. At this point, $r_0$ is left free, but will be set to a specific value after integration over the second coordinate patch. 

Next, in ${\mathcal W}_1$, it is natural to use $v_1$ as the coordinate on the family of $\p^1$'s. Let $v_1=\delta$ be a fixed representative from the class of $\p^1$. We also let $x=\bar u'$ so that the integration over $\Gamma \cong \p^1 \times I$ is sensible: 
\\
$$
\hat W_1(v_1^*)= \int_{\Gamma(v_1^*)}\Omega_1=\int_{v_1=\delta}^{v_1=v_1^*}\int_{\p^1_+}\frac{d{\bar u'}du'}{(1+v_1-|u'|^2)}\frac{dv_1}{v_1}=
$$
\\
$$
\int_{v_1=\delta}^{v_1=v_1^*}\int_{r=0}^{r=\frac{1}{r_0}}\int_{\theta=0}^{\theta=2\pi}\frac{rdrd\theta}{(1+v_1-r^2)}\frac{dv_1}{v_1}=
$$
\\
$$ 
  \int_{v_1=\delta}^{v_1=v_1^*}(log(1+v_1-(\frac{1}{r_0})^2) -log(1+v_1))\frac{dv_1}{v_1}
$$
\\
up to an overall multiplicative factor of $-\pi$. From here, let $r_0=1$; then the above becomes
\\
$$
\int_\delta^{v_1^*} (log(v_1)-log(1+v_1))\frac{dv_1}{v_1}= -\int_\delta^{v_1^*}log\Big(1+\frac{1}{v_1}\Big)\frac{dv_1}{v_1}=
$$
\\

\begin{equation}
\label{opensuper}
\sum_{n>0}\frac{(-(v_1^*)^{-1})^n}{n^2} 
\end{equation}
\\
for the second coordinate patch, up to an additive constant. This is the piece of interest for comparison with the earlier calculations.

\subsection{Equivalence.}

Finally, it will be shown that the result (\ref{closedsuper}) matches (\ref{opensuper}) above.
\begin{thma} 
The functions $W(t_1,t_2), \hat W_1((v_1^*)^{-1})$ agree when $t_2=0$.
\end{thma}
\it Proof. \rm Recall the variable change formulas that were used to exhibit the $B$ model conifold singularity (\ref{changevar}):
\\
\begin{equation}
\label{mirror1}
z_1=\frac{\alpha}{(1+\alpha)^2}, \ z_2=\beta(1+\alpha).
\end{equation}
\\
Note the similarities of this to the exponentiated mirror map (\ref{mirrormap}) 
\\
$$
e^{t_1}=z_1e^{2S}, \ e^{t_2}=z_2e^{-S} 
$$
where
$$
S=\sum_{n_1>0}\frac{(2n_1-1)!}{(n_1!)^2}z_1^{n_1}. 
$$
\\
This is expected, because to see the conifold singularity in the space $Y_{z_1,z_2}$, the volume parameter of the $\p^1$ had to be corrected (via the mirror map) to get $Y_{\alpha,\beta}$. 

To see that the transformations (\ref{mirror1}) and (\ref{mirrormap}) are indeed the same, we can use the formula \cite{CK}
\\
$$
t_1(z)=log(z_1) + 2\sum_{n_1>0}\frac{(2n_1-1)!}{(n_1!)^2}z_1^{n_1} = log \Big(\frac{1-2z_1-\sqrt{1-4z_1}}{2z_1}\Big).
$$
\\
By making the identification $\alpha = e^{t_1}$ and using the above formula, it is easy to show that $e^{t_1}=z_1(1+e^{t_1})^2$. Then, since we also have $e^{t_1}=z_1e^{2S},$ we can conclude that  $1+e^{t_1}=e^{S};$ thus, we find agreement between (\ref{mirror1}) and (\ref{mirrormap}).

Then, from the above, in order to get a match between the period (\ref{closedsuper}) with our open string function (\ref{opensuper}), we must make the same change of coordinates on the function (\ref{closedsuper}) as we did on the space $Y_{z_1,z_2}$; this of course means inserting the mirror map (\ref{mirrormap}) into (\ref{closedsuper}). The result of this is
\\
\begin{equation}
\label{super2}
 W(t_1,t_2)=\sum_{n>0}\frac{e^{nt_2}}{n^2}+\sum_{n>0}\frac{e^{n(t_1+t_2)}}{n^2},
\end{equation}
\\
not including logarithmic terms. For the last step \cite{AV}, recall that there is the identification 
\\
$$
\frac{\partial \mathcal F}{\partial t_2}(t_1,t_2) = W(t_1,t_2).
$$
\\
Then the predicted equivalence of functions is that 
\\
$$
W(t_1,0) = \hat W_1((v_1^*)^{-1}),
$$ 
\\
and this is indeed the case (up to constant and logarithmic terms). \qed

\pagebreak


\begin{thebibliography}{widest-label} 









\bibitem{AV} M. Aganagic and C. Vafa, \it $G_2$ Manifolds, Mirror Symmetry and Geometric Engineering, \rm hep-th/0110171.

\bibitem{V} C. Vafa, \it Superstrings and Topological Strings at Large N, \rm J.Math.Phys. 42 (2001) 2798-2817. hep-th/0008142. 

\bibitem{W} E. Witten, \it Branes And The Dynamics Of QCD, \rm Nucl.Phys. B507 (1997) 658-690. hep-th/9706109. 

\bibitem{H} S. Hosono, \it Central charges, symplectic forms, and hypergeometric series in local mirror symmetry, \rm hep-th/0404043

\bibitem{CKYZ} T.-M. Chiang, A. Klemm, S.-T. Yau and E. Zaslow, \it Local Mirror Symmetry: Calculations and Interpretations, \rm Adv.Theor.Math.Phys. 3 (1999) 495-565. hep-th/9903053

\bibitem{AV2} M. Aganagic and C. Vafa, \it Mirror Symmetry, D-Branes and Counting Holomorphic Discs, \rm  hep-th/0012041. 

\bibitem{MV} M. Mari\~no and C. Vafa, \it Framed knots at large N, \rm hep-th/0108064 .

\bibitem{OV} H. Ooguri and C. Vafa, \it Knot Invariants and Topological Strings, \rm Nucl.Phys. B577 (2000) 419-438. hep-th/9912123.


\bibitem{DFG} D.-E. Diaconescu, B. Florea and A. Grassi, \it Geometric Transitions and Open String Instantons, \rm Adv.Theor.Math.Phys. 6 (2003) 619-642. hep-th/0205234.

\bibitem{CIV} F. Cachazo, K. Intriligator and C. Vafa, \it A Large N Duality via a Geometric Transition, \rm Nucl.Phys. B603 (2001) 3-41. hep-th/0103067.

\bibitem{CKV} F. Cachazo, S. Katz and C. Vafa, \it Geometric Transitions and N=1 Quiver Theories, \rm hep-th/0108120.

\bibitem{AV2} M. Aganagic and C. Vafa, \it Mirror Symmetry and a $G_2$ Flop, \rm JHEP 0305 (2003) 061. hep-th/0105225.


\bibitem{GR} A. Grassi and M. Rossi, \it Large N dualities and transitions in geometry, \rm math.AG/0209044. 

\bibitem{B} V. Batyrev, I. Ciocan-Fontanine, B. Kim and D. van Straten, \it Conifold Transitions and Mirror Symmetry for Calabi-Yau Complete Intersections in Grassmannians, \rm alg-geom/9710022. 

\bibitem{BF} B. Forbes, \it Open string mirror maps from Picard- Fuchs equations on relative cohomology, \rm hep-th/0307167

\bibitem{LM} W. Lerche and P. Mayr, \it{On N=1 Mirror Symmetry for Open Type II Strings},\rm hep-th/0111113  

\bibitem{CK} D. Cox and S. Katz, \it Mirror symmetry and algebraic geometry, \rm Mathematical Surveys and Monographs, 68. American Mathematical Society, Providence, RI, 1999 

\bibitem{LLZ} C.-C. Liu, K. Liu and J. Zhou, \it A Proof of a Conjecture of Marino-Vafa on Hodge Integrals, \rm J. Differential Geom. 65 (2003), no. 2, 289--340. math.AG/0306434








\end{thebibliography}
\end{document}